# ML-PCM : Machine Learning Technique for Write Optimization in Phase Change Memory (PCM)

Mahek Desai[1], Rowena Quinn[2], and Marjan Asadinia[1]

[1] California State University, Northridge, USA
mahek-trushit.desai.849@my.csun.edu
marjan.asadinia@csun.edu
[2] Illinois Central College, Peoria, USA
rowenaquinn1@gmail.com

**Abstract.** As transistor-based memory technologies like dynamic random access memory (DRAM) approach their scalability limits, the need to explore alternative storage solutions becomes increasingly urgent. Phase-change memory (PCM) has gained attention as a promising option due to its scalability, fast access speeds, and zero leakage power compared to conventional memory systems. However, despite these advantages, PCM faces several challenges that impede its broader adoption, particularly its limited lifespan due to material degradation during write operations, as well as the high energy demands of these processes. For PCM to become a viable storage alternative, enhancing its endurance and reducing the energy required for write operations are essential. This paper proposes the use of a neural network (NN) model to predict critical parameters such as write latency, energy consumption, and endurance by monitoring real-time operating conditions and device characteristics. These predictions are key to improving PCM performance and identifying optimal write settings, making PCM a more practical and efficient option for data storage in applications with frequent write operations. Our approach leads to significant improvements, with NN predictions achieving a Mean Absolute Percentage Error (MAPE) of 0.0073% for endurance, 0.23% for total write latency, and 4.92% for total write energy.

## 1 Introduction

With DRAM and other transistor-based memory technologies reaching their scalability constraints, the search for new storage alternatives is becoming more critical. Phase Change Memory (PCM) has emerged as a strong candidate, leveraging chalcogenide materials like $Ge_2Sb_2Te_5$ (GST) [1] to toggle between low-resistance crystalline and high-resistance amorphous states for data storage. PCM presents several advantages, including high scalability, non-volatility, low leakage power, and competitive read latency, making it a promising contender for future main memory systems [2] – [5]. PCM operates primarily through two processes: SET and RESET operations [6]. The SET operation involves heating the material below its melting point and then gradually cooling it to achieve a crystalline state, which represents a "1" in memory. Conversely, the RESET operation heats the material above its melting point and then rapidly cools it, creating an amorphous state, which represents a "0". These operations demand significant electrical power due to the necessary heating and cooling.

However, PCM is not without challenges, particularly regarding write endurance. Typically, PCM cells endure only about $10^7$ to $10^9$ write cycles before degradation, leading to potential data retention issues such as "stuck-at" faults where cells remain permanently in a specific state [3, 5]. This limitation impacts reliability and longevity, especially in applications with frequent write operations. Additionally, PCM write operations are more energy-intensive compared to DRAM, contributing to higher power consumption. The latency of write operations also poses a concern, as the material requires cooling time between operations, which can affect overall system performance.

To address these challenges, a novel approach involves integrating neural network (NN) models, with potential future enhancements. The NN model is designed to predict key metrics such as write energy, latency, and endurance based on parameters like voltage, current, and pulse duration under varying conditions, including voltage fluctuations. In parallel. This innovative application of machine learning represents a new approach to overcoming critical limitations in PCM technology.

The remainder of this paper is structured as follows: Section 2 explores the related works. Section 3 presents the proposed method, detailing the integration of the NN model for optimizing write parameters. Section 4 discusses the evaluation of our results, illustrating the simulator we will be implementing to generate our data and compare our results, and highlighting the accuracy of our NN. Section 5 dives deep into the future works. Finally, section 6 concludes the paper, summarizing our contributions and the potential impact of our proposed methods on the development of PCM as a next-generation memory technology.

## 2 Related Works

In recent years, several studies have explored the optimization of write cycles and power efficiency in Non-Volatile Memory (NVM), especially Phase Change Memory (PCM) using various machine learning techniques.

Lim et al.[7] introduced a novel workload-aware and optimized write cycle management system for NVRAM. Their approach focused on improving endurance and performance by dynamically adjusting write strategies based on workload characteristics, which helps to distribute write operations more evenly across memory cells, thereby prolonging the lifespan of the memory and en-hancing overall performance. Another significant contribution presented in [8] demonstrates the effectiveness of power-aware reinforcement learning in managing memory power consumption. Their method leverages reinforcement learning algorithms to predict and adjust power states dynamically, thus significantly reducing energy usage while maintaining performance. The proposed system learns from historical power consumption patterns and optimizes the power states accordingly, leading to substantial energy savings without compromising the performance of the memory system.

In [9], authors proposed a deep neural network model designed for accurate and efficient performance prediction in memory systems. Their model emphasizes the ability to handle complex memory access patterns and predict performance metrics with high accuracy. By utilizing a deep neural network, the model can capture intricate relationships within the memory access patterns, enabling more precise performance predictions and, consequently, better optimization strategies for memory management. The work by Chen et al.[10] highlighted the potential of distributed reinforcement learning for optimizing power consumption in large-scale memory systems. Their approach involves deploying reinforcement learning agents across a distributed memory architecture to collaboratively optimize power usage. This method not only improves power efficiency but also enhances computational performance by balancing the load and reducing bottlenecks in the memory system. The distributed nature of the solution ensures scalability and robustness, making it suitable for large-scale implementations.

Additionally, in [11], the authors present a technique to eliminate redundant writes, reducing energy use by up to 50% and doubling endurance. Their method shows minimal performance impact, enhancing PCM's reliability and efficiency for future systems. The modular reinforcement learning approach presented by Kim et al. [12] provides a flexible framework for integrating various reinforcement learning models to manage different aspects of memory systems. This approach allows for the simultaneous optimization of multiple parameters, such as power consumption, latency, and endurance, by utilizing specialized reinforcement learning agents for each task.

The self-optimizing memory controller in [13] demonstrates that by dynamically adjusting scheduling decisions based on past behaviors, memory access latencies can be significantly reduced, thereby improving overall system performance. For modern memory systems,[14] introduces Memory Cocktail Therapy (MCT), which leverages machine learning techniques and architectural insights to optimize memory system performance. By employing gradient boosting and quadratic lasso models, MCT dynamically selects the best memory management policy based on runtime conditions. This approach has been shown to significantly improve instructions per cycle (IPC), extend memory system lifetime, and reduce overall energy consumption compared to static policies, making it a highly effective method for modern memory systems.

Furthermore, reinforcement learning-based methods have also shown promise in the domain of adaptive caching and refresh optimization. These methods use reinforcement learning algorithms to make real-time decisions about cache management and refresh operations, learning from access patterns to minimize refreshes and improve cache hit rates. This adaptive approach allows for a more efficient and responsive memory system that can better handle the dynamic nature of workload demands [15], [16].

## 3 Proposed Method

In this section, we provide a detailed explanation of our proposed ML-PCM approach. We start by discussing the data collection process for the write operation, followed by an overview of the data preprocessing techniques used to clean and prepare the data for training machine learning models.

### 3.1 Parameter Generation

To optimize write energy and latency, we generated parameters for the NVMain simulator [17, 18], focusing on various voltage levels and pulse durations for both set and reset operations, as well as different read:write ratios. The set operation parameters included discrete voltage values of 1.5 V, 2.0 V, and 2.5 V, and pulse durations of 150 ns, 155 ns, and 160 ns. For the reset parameters, the voltage was varied discretely with values 2.5 V, 3.0 V, and 3.5 V, and the pulse durations considered were 100 ns, 105 ns, and 110 ns. These ranges were selected because we obtained consistent results across different ranges, demonstrating robustness in our findings.

We also generated trace files as inputs to the NVMain simulator, varying the read:write ratios, including 9 : 1, 8 : 2, 7 : 3, and others. Specifically, we created 20 trace files where read operations outnumber write operations ($R > W$), 20 files for balanced operations ($R = W$), and 20 files where write operations exceed read operations ($R < W$). A comprehensive dataset was built by combining and permuting all these variables to cover a broad range of operational conditions.

The next section will outline our methodology, with the flow between each section visualized in Figure 1.

### 3.2 Simulation Process

The simulation process in NVMain starts by reading the first row from the dataset file, which includes parameter combinations. These parameters, such as set voltage, reset voltage, set pulse duration, and reset pulse duration, are then updated in NVMain according to the values read. NVMain is executed using the corresponding trace file for that parameter set, and the simulation runs until 100,000 operations are completed, equating to approximately 7.5 million simulation cycles. The results are stored in a designated folder, and the process is repeated for each row in the dataset until all rows are processed.

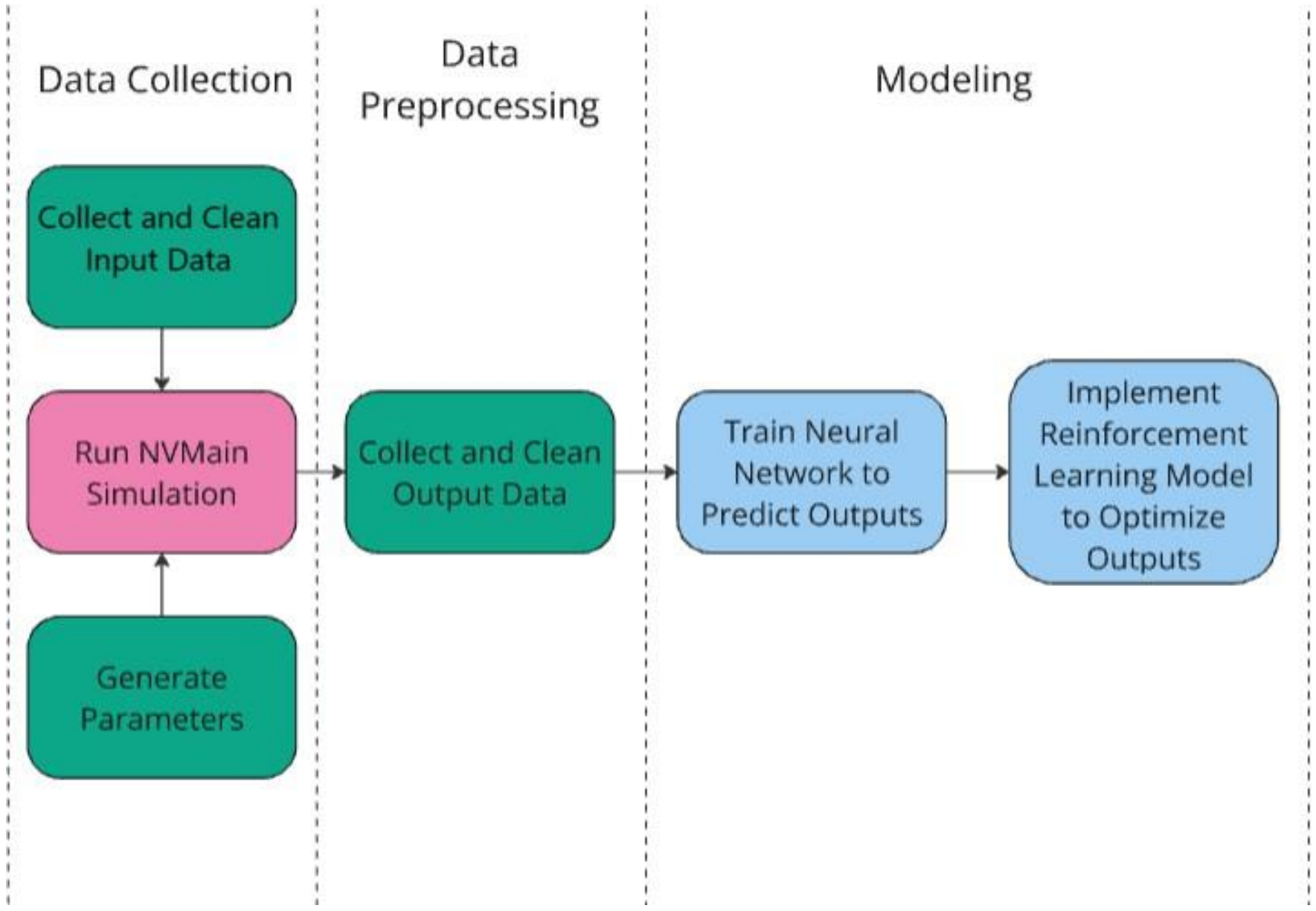

**Fig. 1.** ML-PCM Overview.

### 3.3 Data Preprocessing

Post-simulation, we gathered the raw output data corresponding to the input values in NVMain. This raw data was cleaned to extract key metrics such as write energy, total energy, write latency, total latency, and endurance.

To prepare the data for the predictive model, several preprocessing steps were undertaken. Discrete input variables were One-Hot Encoded to eliminate ordinality during model training. After preprocessing, we had 4,860 rows of data, which were split into training, testing, and validation sets. 60% of the data was allocated for training, 20% for testing, and 20% for validation, ensuring the model's robustness and generalization to new data, avoiding overfitting on the simulation data.

### 3.4 NN Model

ML-PCM leverages neural networks (NN) for prediction due to their superior capabilities in addressing the complexities of phase-change memory (PCM) systems. Neural networks excel in modeling complex, non-linear relationships and capturing intricate patterns in data, which is crucial for accurate write energy predictions. They outperform alternatives like decision trees, linear regression and polynomial regression which struggle with non-linearity, and support vector machines (SVMs), which lack the depth required for nuanced predictions.

After data collection from NVMain, we separated the data into inputs (features) and outputs (targets). We selected voltage and pulse duration for set and reset operations, along with the total reads and writes from a trace file as features, while the targets were total write energy, total write latency, and endurance. The model was developed using a Multi-Layer Perceptron (MLP) architecture with these input features and output targets.

**Table 1.** Input Features with Linked Output Features.

| Input Features | Output Features |
|---|---|
| SET/RESET Voltage<br>SET/RESET Pulse Duration<br>Read:Write Ratio | Total Write Energy |
| SET/RESET Pulse Duration<br>Read:Write Ratio | Total Write Latency |
| Read: Write Ratio | Endurance per Bank |

The MLP Neural Network (NN) was structured as a multi-output regression model using Keras's Functional API [19] and optimized using the Keras Tuner [19]. By analyzing NVMain's method of calculating write energy, latency, and endurance per bank, we isolated the relevant inputs to train the model for the respective outputs.

Table 1 outlines the output features along with their corresponding input features.

This model took multiple features and separated them into 6, 5, and 5 dense layers to predict the corresponding targets (write energy, write latency, and endurance per bank).

Hyperparameter tuning was performed using the Keras [19] tuner for the following aspects: the number of hidden layers, the number of neurons per layer, the loss function, L1 and L2 values for the kernel regularizer, batch size, and the optimizer. These hyperparameters, which are adjustable during training, were tuned by running multiple combinations until the combination with the lowest mean absolute error (MAE) and validation loss was identified. The model settled on the Adam optimizer, Huber loss function, and a batch size of 160. The resulting differences in model structure when separating into different dense layers are shown in Table 2, and a diagram of the MLP NN structure is depicted in Figure 2.

**Table 2.** Neural Network Architecture For Each Output.

| Model Output | Hidden Layers | Neurons per Layer | Kernel Regularization L1, L2 Weights |
|---|---|---|---|
| Write Energy | 6 | 28,28,14,6,6,16 | 0.001,0.001 |
| Write Latency | 5 | 30,14,24,16,12 | 0.01,0.001 |
| Endurance | 5 | 30,14,24,16,8 | 0.01,0.001 |

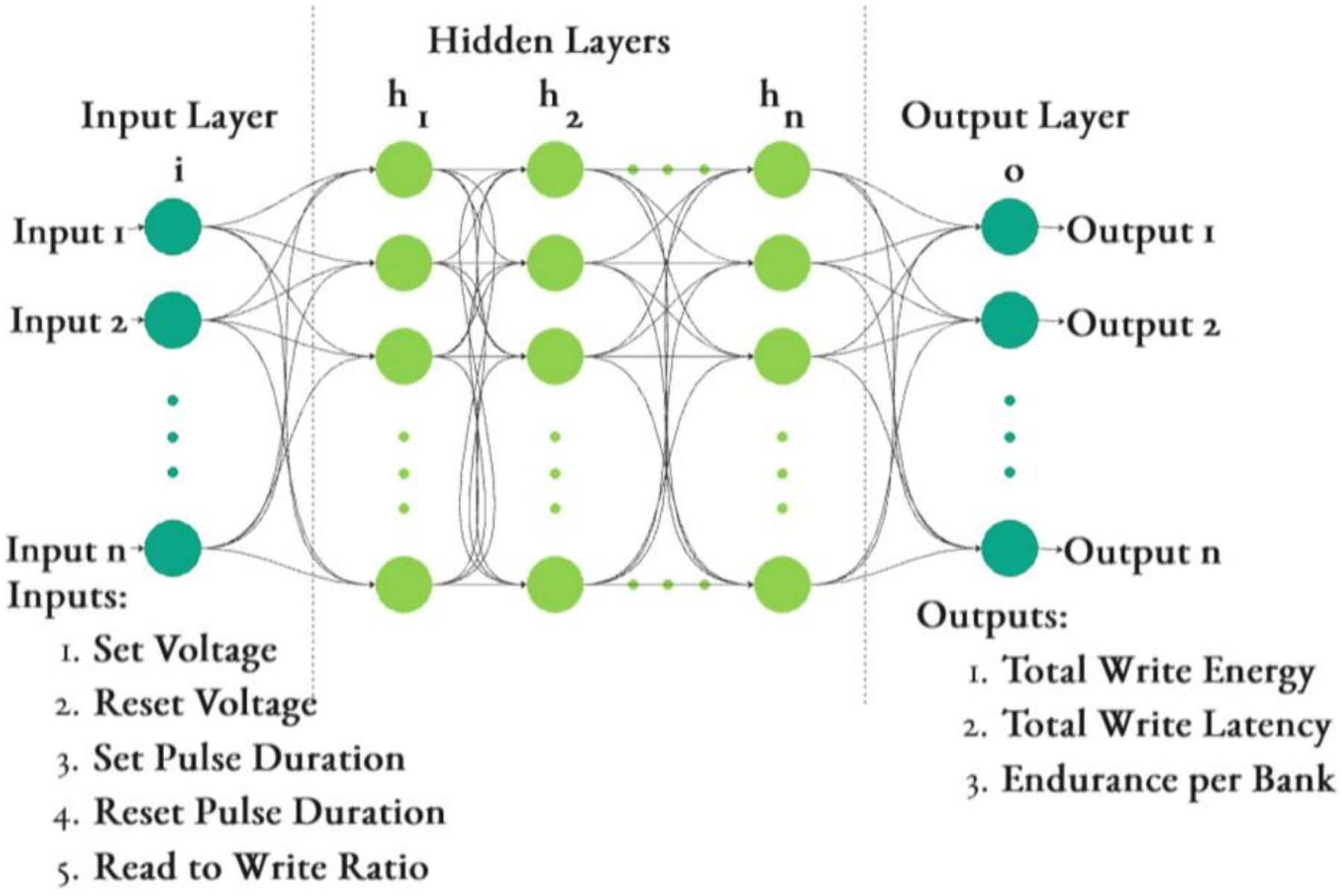

**Fig. 2.** Multi-Layer Perceptron Neural Network Architecture.

During training, two callbacks were used to optimize training performance and minimize validation loss. The learning rate for the Adam optimizer was initially set to 0.001 and reduced by 0.1 whenever the validation loss plateaued for three epochs. Training was stopped if the validation loss did not improve for four consecutive epochs, preventing overfitting.

## 4 Evaluation

In this section, we describe our simulation environment, methodology, and present the results of evaluating the proposed method against a baseline PCM configuration.

### 4.1 Simulation Environment

Our proposed method was implemented in NVMain [17, 18], a comprehensive, cycle-accurate memory simulator capable of modeling both traditional DRAM and emerging non-volatile memory (NVM) technologies, including phase change memory (PCM). NVMain provides detailed simulations of memory performance, energy consumption, and NVM-specific characteristics such as limited write endurance and multi-level cells. Additionally, NVMain supports hybrid memory architectures, fine-grained bank/subarray-level parallelism, and allows for the customization of memory controllers and address mapping schemes. For PCM simulations, NVMain effectively captures critical attributes like asymmetric read/write latencies, write energy, and cell endurance, making it well-suited for evaluating PCM optimization strategies [17, 18]. NVmain's simulation process is as shown in Figure 3.

The simulation configuration is detailed in Table 3. The setup consists of a 20nm, 1.8V, 8Gb PRAM with a 40MB/s programming bandwidth. Other parameters include clock frequency, bus width, CPU frequency, and more, which were maintained in their default settings within NVMain to accurately simulate PCM. These parameters are vital for thoroughly evaluating memory system performance and energy consumption under various scenarios.

**Table 3.** Simulation Configuration.

| Parameter | Value |
|---|---|
| **Memory Specifications** | |
| Technology node | 20nm |
| Operating voltage | 1.8V |
| Device capacity | 8Gb |
| Program bandwidth | 40MB/s |
| **Interface Specifications** | |
| Clock frequency in MHz | 400 |
| Bus width in bits | 64 |
| Number of bits per device | 8 |
| CPU frequency in MHz | 2000 |
| **MLC Parameters** | |
| Number of MLC levels | 2 |
| **Memory Controller Parameters** | |
| Memory controller type | FRFCFS |
| Address mapping scheme | R:RK:BK:CH |
| Read queue size | 32 |
| Write queue size | 32 |
| **Endurance Model Parameters** | |
| Endurance model type | BitModel |
| Endurance distribution type | Normal |
| Endurance distribution mean | 1000000 |
| Endurance distribution variance | 100000 |

## 4.2 Evaluation of NN Models

Given that the NN's objective was regression, measuring accuracy involved more than simply calculating the proportion of correct predictions. Thus, we employed Mean Absolute Percentage Error (MAPE) to assess the model's performance. MAPE is calculated as the mean of the absolute differences between the actual and predicted values, divided by the actual values. This metric gives a percentage indicating the closeness of the predicted values to the actual values, with lower percentages indicating a better fit.

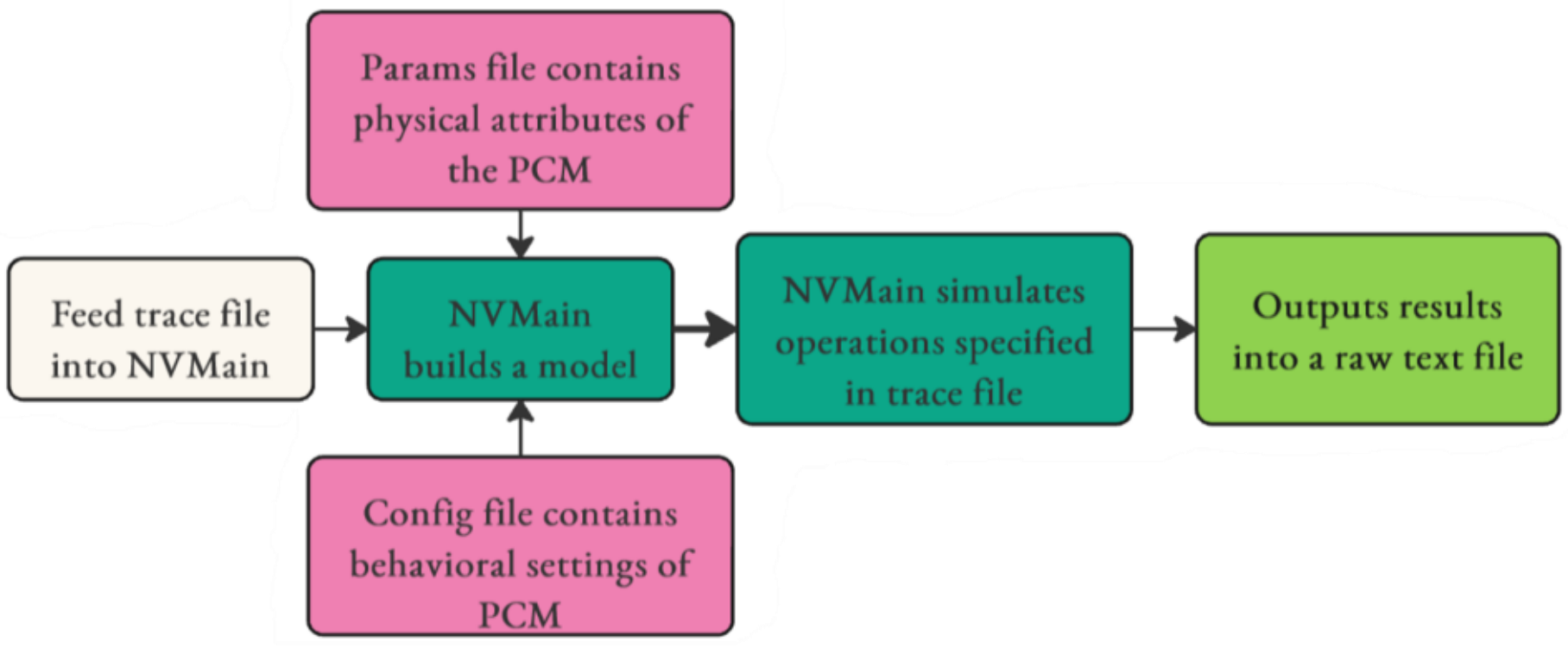

**Fig. 3.** NVMain Simulation Process.

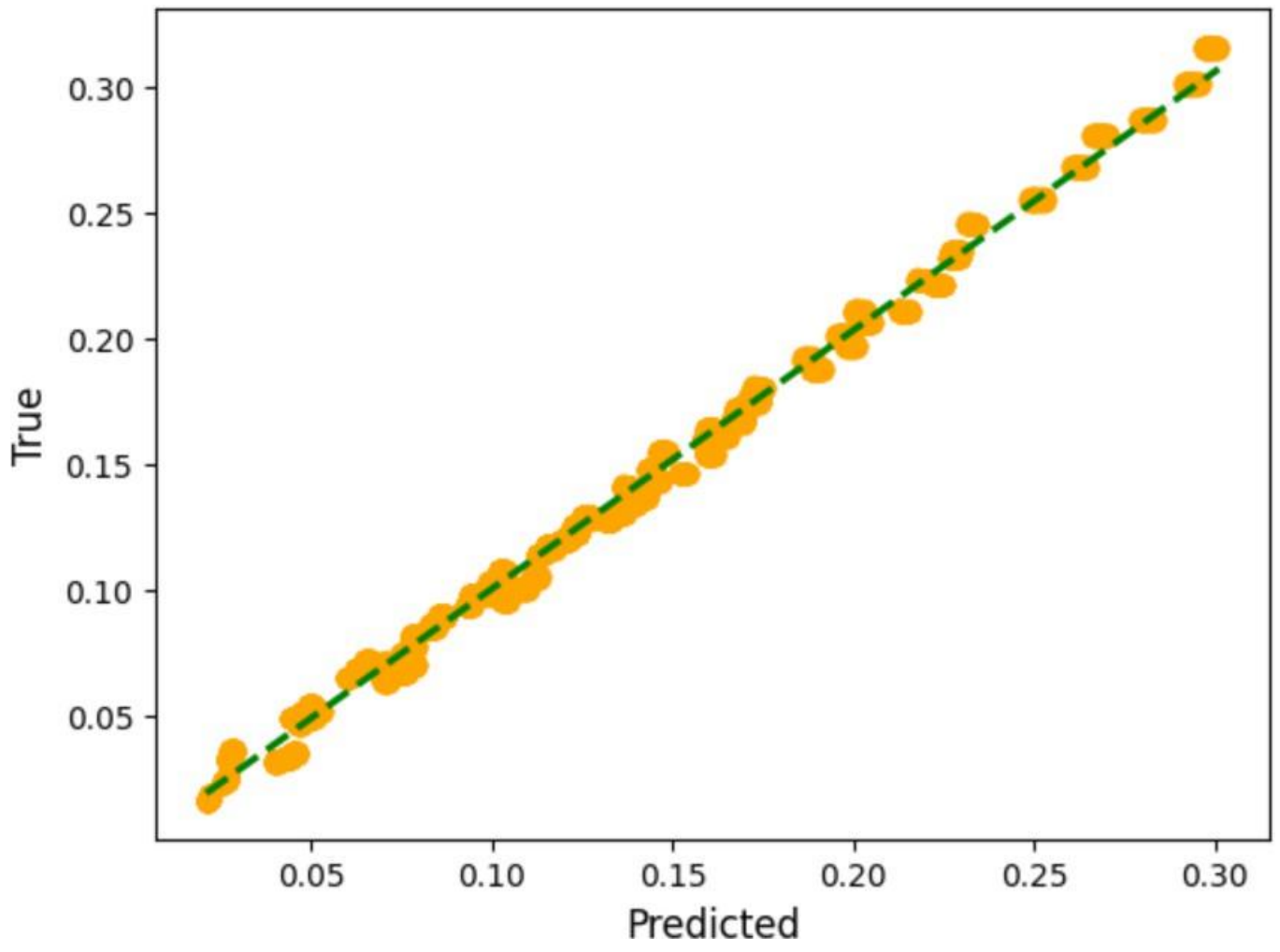

**Fig. 4.** Regression of Predicted vs. Actual Write Energy.

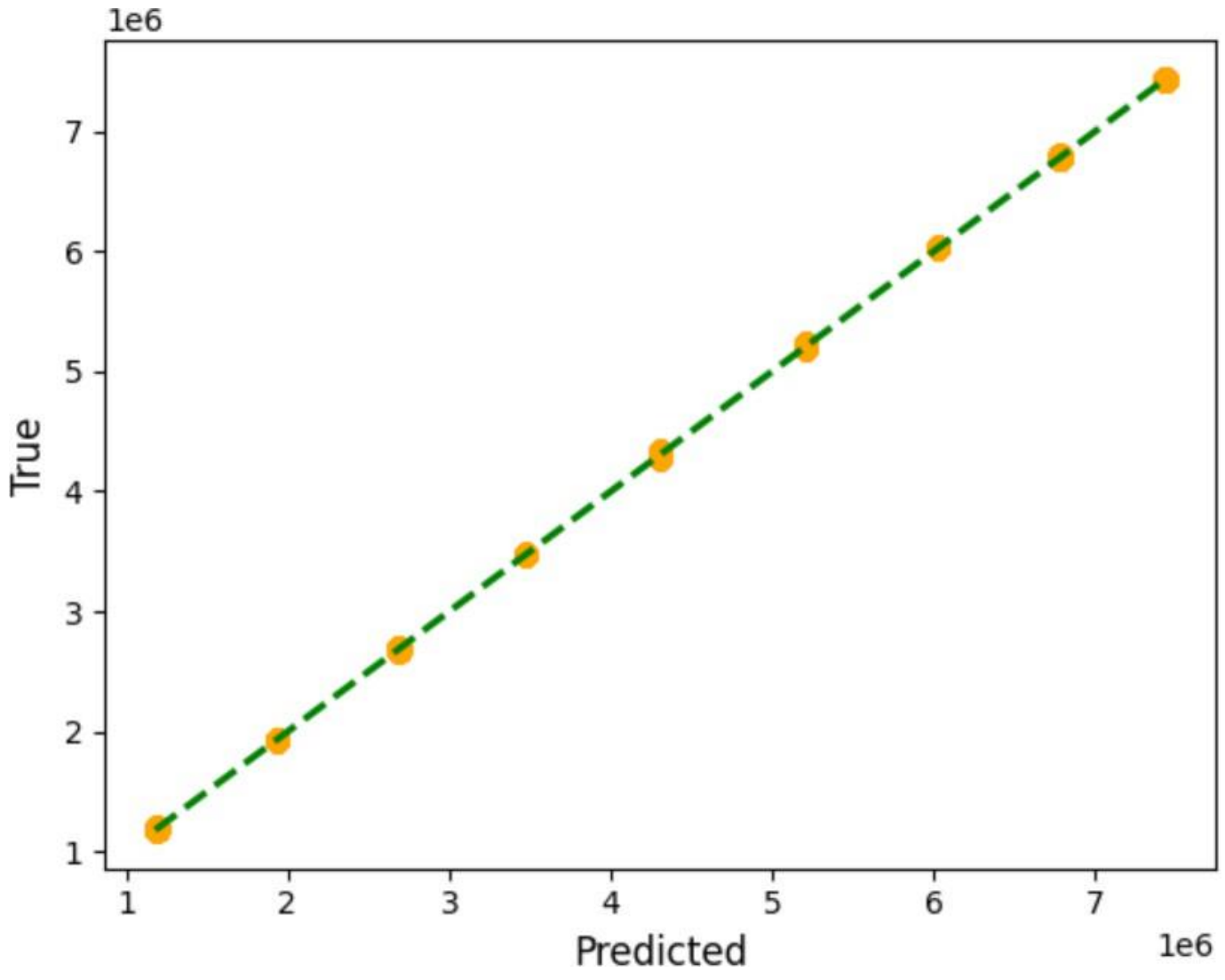

**Fig. 5.** Regression of Predicted vs. Actual Write Latency.

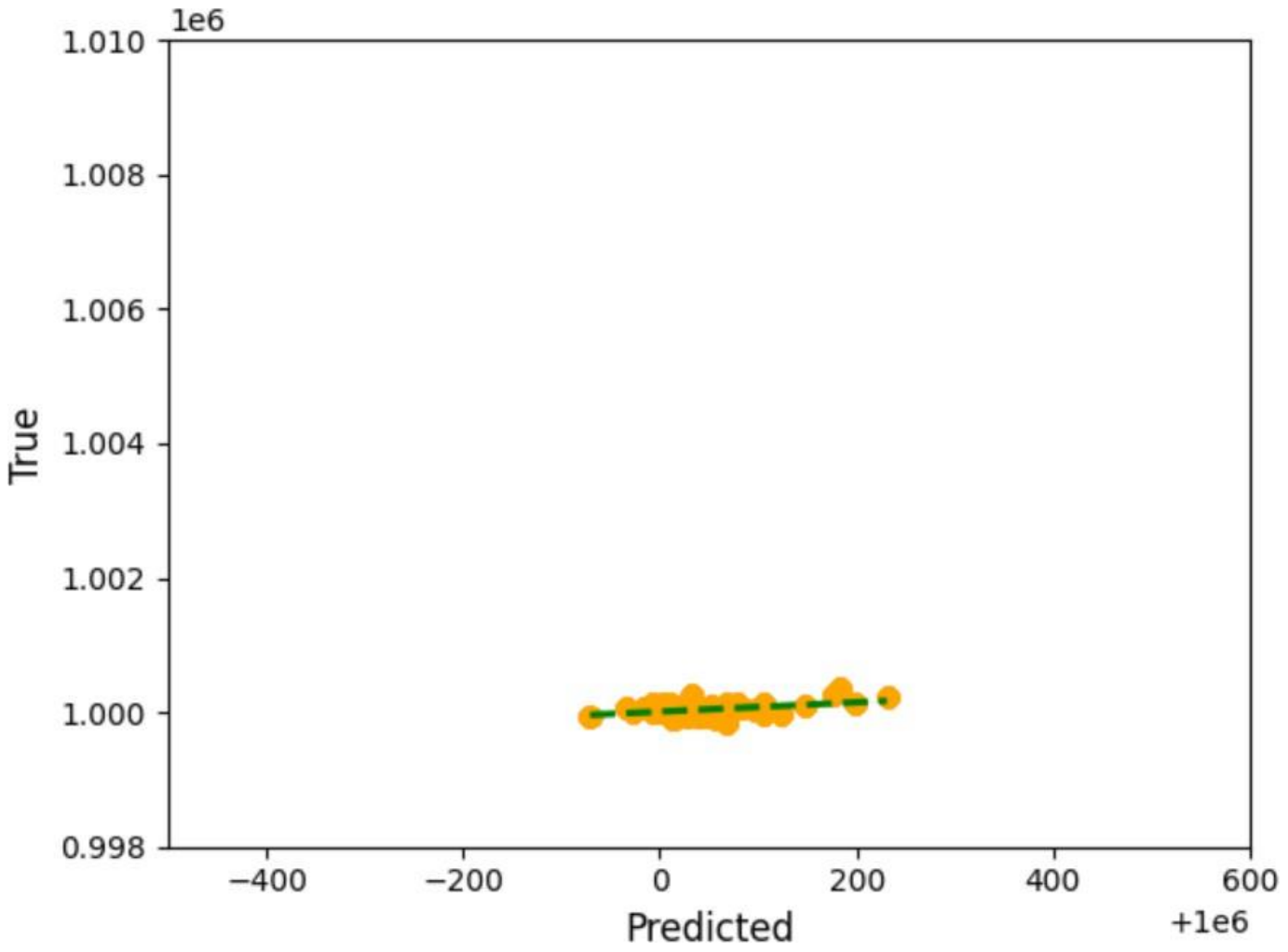

**Fig. 6.** Regression of Predicted vs. Actual Endurance.

During the training phase, as discussed in the Section 3.3, we divided the data into separate sets for training, testing, and validation. Specifically, 60% of the data was allocated for training, while 20% was set aside for both testing and validation. Initially, we evaluated how well the model fit the training data, followed by an assessment of how well it generalizes to new data, ensuring the model's adaptability to various datasets. This required evaluating both the training loss MAPE (model fit to training data) and validation loss MAPE (model fit to unseen data).

**Table 4.** MAPE For Each Output.

| Output | MAPE |
|---|---|
| Endurance | 0.0073% |
| Total Write Latency | 0.23% |
| Total Write Energy | 4.92% |

The trends for loss in both the training and validation sets showed a significant initial decrease, followed by gradual reductions, thanks to the learning rate suppression callback implemented during model fitting. Since the validation loss closely mirrored the training loss, we inferred that the model did not overfit to the training data.

After ensuring the model was not overfitted, we proceeded to evaluate the NN's accuracy. We utilized MAPE to assess the testing data, achieving the results shown in Table 4.

Subsequently, we generated regression plots, shown in Figures 4, 5, and 6, which depict the deviation of predicted values from actual values.

Overall, these results demonstrate the effectiveness of the model in predicting key performance metrics, such as endurance, write latency, and energy consumption, with impressive accuracy. This provides valuable stepping stone towards optimizing Phase Change Memory (PCM), enabling more low energy and faster write operations while maintaining endurance levels, contributing to the advancement of PCM technology as main memory.

## 5 Future Works

In future research, we plan to enhance the proposed method by incorporating the following aspect:

**Integration of Temperature Model** Incorporate a temperature model that accounts for variations in $Ge_2Sb_2Te_5$ (GST) which affect resistance and optimal operational parameters.

Recent studies on chalcogenide glass materials (GST) used in PCM have made progress in modeling the relationship between ambient temperature and the energy required for a RESET operation. According to [20], a linear relationship was identified between power density and ambient temperature, expressed by the equation 1

$$P = g - hT_{\text{ambient}} \tag{1}$$

where g = (32.9 ± 0.1) MW/cm2 and h = (0.04 ± 0.0003) MW/cm2K [20]. This equation allows us to determine the energy required for a RESET operation based on ambient temperature. By incorporating these temperature-dependent discrete input variables into the NN model, we can optimize the total write energy and latency, adapting the model to account for system temperature variations.

## 6 Conclusion

In conclusion, this paper introduces an innovative approach to optimizing PCM write energy consumption and performance using neural networks (NN). By predicting key write parameters based on real-time operational conditions and device characteristics, our proposed method aims to reduce write energy consumption and latency while improving PCM endurance. Initial findings from our evaluation involved comprehensive parameter generation, simulation processes using NVMain, and modeling a predictive NN model, which achieved a Mean Absolute Percentage Error (MAPE) of 0.0073% for endurance, 0.23% for total write latency, and 4.92% for total write energy. Future works include integrating a temperature model to better predict write parameters.

## 7 Acknowledgment

This work was supported by the National Science Foundation (NSF) under Grant Nos. [CNS-2244391] and [2318553]. We gratefully acknowledge the NSF for their financial support through both grants and for providing the resources necessary to conduct this research. The views expressed in this paper are those of the authors and do not necessarily reflect the views of the NSF.

## References

1. A. Ehrmann, T. Blachowicz, G. Ehrmann, and T. Grethe, "Recent developments in phase-change memory," Applied Research, Jun. 2022, doi: https://doi.org/10.1002/appl.202200024.
2. R. Azevedo, J. D. Davis, K. Strauss, P. Gopalan, M. Manasse, and S. Yekhanin, "Zombie memory: Extending memory lifetime by reviving dead blocks," in *Proceedings of the International Symposium on Computer Architecture* (ISCA), 2013.

3. H. Luo et al., “Write Energy Reduction for PCM via Pumping Efficiency Improvement,” *ACM Transactions on Storage*, vol. 14, no. 3, pp. 1–21, Aug. 2018.
4. J. Fan, S. Jiang, J. Shu, Y. Zhang, and W. Zhen, “Aegis: Partitioning data block for efficient recovery of stuck-at-faults in phase change memory,” in *Proceedings of the 46th Annual IEEE/ACM International Symposium on Microarchitecture*, pp. 433–444, ACM, 2013.
5. P. Zhou, B. Zhao, J. Yang, and Y. Zhang, “Throughput Enhancement for Phase Change Memories,” *IEEE Transactions on Computers*, vol. 63, no. 8, pp. 2080–2093, 2014.
6. Zhou, Y., Zhang, W., Ma, E. et al. Device-scale atomistic modelling of phase-change memory materials. Nat Electron 6, 746–754 (2023). https://doi.org/10.1038/s41928-023-01030-x
7. J. P. Shri Tharanyaa, D. Sharmila, and R. Saravana Kumar, “A Novel Workload-Aware and Optimized Write Cycles in NVRAM,” Computers, Materials & Continua, vol. 71, no. 2, pp. 2667–2681, 2022, doi: https://doi.org/10.32604/cmc.2022.019889.
8. K. Fatemi and Masoud, “A Power-Aware Reinforcement Learning Technique for Memory Allocation in Real-time Embedded Systems,” Dec. 22AD, Available: https://prism.ucalgary.ca.
9. M. Shi, P. Mo, and J. Liu, “Deep Neural Network for Accurate and Efficient Atomistic Modeling of Phase Change Memory,” IEEE Electron Device Letters, vol. 41, no. 3, pp. 365–368, Mar. 2020, doi: https://doi.org/10.1109/led.2020.2964779.
10. Z. Chen and D. Marculescu, “Distributed Reinforcement Learning for Power Limited Many-Core System Performance Optimization,” Design, Automation & Test in Europe Conference & Exhibition (DATE), 2015, Jan. 2015, doi: https://doi.org/10.7873/date.2015.0992.
11. S. Song, A. Das, O. Mutlu, and N. Kandasamy, “Improving phase change memory performance with data content aware access,” arXiv (Cornell University), Jun. 2020, doi: https://doi.org/10.1145/3381898.3397210.
12. Z. Wang et al., “Modular Reinforcement Learning for self-adaptive Energy Efficiency Optimization in Multicore System,” Jan. 2017, doi: https://doi.org/10.1109/aspdac.2017.7858403.
13. E. Ipek, O. Mutlu, J. F. Martínez, and R. Caruana, “Self-Optimizing Memory Controllers: A Reinforcement Learning Approach,” 2008 International Symposium on Computer Architecture, Jun. 2008, doi: https://doi.org/10.1109/isca.2008.21.
14. Z. Deng, L. Zhang, N. Mishra, H. Ho!mann, and F. T. Chong, “Memory Cocktail Therapy: A General Learning-Based Framework to Optimize Dynamic Tradeoffs in NVMs,” 2017, doi: https://doi.org/10.1145/3123939.3124548.
15. S. Suman and H. K. Kapoor, “Reinforcement Learning Based Refresh Optimized Volatile STT-RAM Cache,” Jul. 2020, doi: https://doi.org/10.1109/isvlsi49217.2020.00048.
16. A. Sadeghi, F. Sheikholeslami, A. G. Marques, and G. B. Giannakis, “Reinforcement Learning for Adaptive Caching With Dynamic Storage Pricing,” IEEE Journal on Selected Areas in Communications, vol. 37, no. 10, pp. 2267–2281, Oct. 2019, doi: https://doi.org/10.1109/jsac.2019.2933780.
17. M. Poremba and Y. Xie, "NVMain: An Architectural-Level Main Memory Simulator for Emerging Non-volatile Memories," in *2012 IEEE Computer Society Annual Symposium on VLSI*, Amherst, MA, 2012, pp. 392-397, doi: 10.1109/ISVLSI.2012.82.

18. M. Poremba, T. Zhang and Y. Xie, "NVMain 2.0: A User-Friendly Memory Simulator to Model (Non-)Volatile Memory Systems," *IEEE Computer Architecture Letters*, vol. 14, no. 2, pp. 140-143, July-Dec. 2015, doi: 10.1109/LCA.2015.2402435.
19. Chollet, F., “keras,” GitHub, Jul. 07, 2021. https://github.com/fchollet/keras.
20. Z. Chen and D. Marculescu, “Distributed Reinforcement Learning for Power Limited Many-Core System Performance Optimization,” Design, Automation & Test in Europe Conference & Exhibition (DATE), 2015, Jan. 2015, doi: https://doi.org/10.7873/date.2015.0992.